# Disentangling ferroelectric wall dynamics and identification of pinning mechanisms via deep learning

Yongtao Liu,[1] Roger Proksch,[2] Chun Yin Wong,[3] Maxim Ziatdinov,[1,4] and Sergei V. Kalinin[1]

[1] Center for Nanophase Materials Sciences, Oak Ridge National Laboratory, Oak Ridge, TN 37831, USA

[2] Asylum Research, An Oxford Instruments Company, Santa Barbara, CA 93117, USA

[3] The Bredesen Center, The University of Tennessee, Knoxville, TN 37996, USA

[4] Computational Sciences and Engineering Division, Oak Ridge National Laboratory, Oak Ridge, TN 37831, USA

Field-induced domain wall dynamics in ferroelectric materials underpins multiple applications ranging from actuators to information technology devices and necessitates a quantitative description of the associated mechanisms including giant electromechanical couplings, controlled non-linearities, or low coercive voltages. While the advances in dynamic Piezoresponse Force Microscopy measurements over the last two decades have rendered visualization of polarization dynamics relatively straightforward, the associated insights into the local mechanisms have been elusive. Here we explore the domain dynamics in model polycrystalline materials using a workflow combining deep learning-based segmentation of the domain structures with non-linear dimensionality reduction using multilayer rotationally-invariant autoencoders (rVAE). The former allows unambiguous identification and classification of the ferroelectric and ferroelastic domain walls. The rVAE discover the latent representations of the domain wall geometries and their dynamics, thus providing insight into the intrinsic mechanisms of polarization switching, that can further be compared to simple physical models. The rVAE disentangles the factors affecting the pinning efficiency of ferroelectric walls, offering insights into the correlation of ferroelastic wall distribution and ferroelectric wall pinning.

Ferroelectric materials are key in applications ranging from electromechanical sensors and actuators, electrooptical filters and modulators, capacitors, gates, and transport-based information technology devices.[1-5] Functional behavior of these materials in these applications is inherently linked to the polarization dynamics, including processes such as the nucleation of new domains,[6-9] motion and pinning of domain walls,[7, 10-13] and polarization rotations[9, 14]. Consequently, exploring the fundamental mechanisms of the domain wall motion in ferroelectric materials has become one of the long-standing problems in modern physics.[15-18]

The unique aspect of ferroelectric materials compared to other materials classes is the presence of significant lattice deformation and built-in dipole. The former results in the strong coupling between the non-180 (ferroelastic) domain walls and external strain fields. This results both in enhanced electromechanical responses due to ferroelastic wall dynamics and polarization rotations, and leads to strong sensitivity of the latter to external strain fields and boundary conditions.[19-21] The polar nature of the ferroelectric materials imposes the formation of bound charge at the walls with the discontinuity of polarization, resulting in characteristic instabilities of charged walls[22] and strong coupling between wall behavior and semiconducting[23] and electrochemical phenomena at surfaces and interfaces.[24, 25] Notably that while ideal single crystals can form the low energy ground states with a periodic domain wall structure and minimal electrostatic and strain energy, realistic materials evolve complex non-equilibrium domain wall structures due to complex histories, non-local effects due to presence of grains and surfaces, and defects and disorder.

Over the first half a century of physics of ferroelectrics, exploration of the domain wall dynamics was preponderantly based on theoretical models and macroscopic measurements on samples with multiple domain walls, with the additional insight from relatively low-resolution optical studies using polarized light or chemically etched/decorated samples. The breakthrough in exploration of ferroelectric domain wall dynamics was achieved with the introduction of Piezoresponse Force Microscopy by Kolosov,[26] Gruverman,[27, 28] Franke,[29, 30] and Takata.[31] Immediately after the invention PFM has become the technique of choice for exploration of ferroelectric materials and multiple studies of static domain structures, their evolution during phase transitions and in applied electric fields. For the latter, particularly broad sets of studies were enabled since the PFM tip can be employed as a probe of polarization dynamics induced by lateral electrodes deposited on the surface,[32] image polarization dynamics through the top electrode in capacitor structures,[33, 34] or used to switch polarization directly.[35] Using these approaches, multiple studies of polarization dynamics in uni- and multi-axial ferroelectrics and devices have been reported.[36-41]

The proliferation of the PFM based polarization studies and observations of domain wall dynamics, in turn, necessitates the development of analytical tools connecting the observational data to the fundamental physical mechanisms. Until now, the vast majority of these analyses were based on the mechanistic analysis of wall velocities as a function of the field or local curvature.[42, 43] In several cases, wall geometry was explored via fractal geometry based descriptors,[44-47] allowing for identification of the theoretically-explored universality classes. However, the full range of phenomena encoded in the domain wall geometries generally remained unexplored.

Similarly, quantification of qualitatively established mechanisms such as wall rotations, nucleation, wall-defect interactions from experimental data remained largely unknown.

Here we explore the time-dependent domain wall dynamics in ferroelectric materials via latent representation of the time-dependent data.[48-50] We propose the feature-engineering approach that effectively encodes time-dependent wall dynamics and demonstrate the applicability of rotationally invariant variational autoencoder to establish the salient features of the domain wall dynamics. The individual elements of this workflow as well as the integrated approach *per se* can be universally applied to the multitude of PFM applications.

Our goal is to develop a universal approach (Figure 1) to analyze the bias- and time-dependent phenomena in ferroelectrics from observations of domain wall dynamics. The starting point in such analysis is PFM images obtained as a function of time and bias. Such images contain features pertaining to the ferroelectric domain walls, as well as contrast variations associated with crosstalk with topography, structural defects, *etc*. Hence, the first step of the analysis workflow is the development of the deep convolutional neural networks (DCNN, Figure 1b) for image segmentation, *i.e.,* pixel-level classification of the image as a ferroelectric or ferroelastic wall or non-domain wall. This step can naturally be combined with classification, *i.e.,* ferroelectric and ferroelastic walls can be considered as the same or different classes. This DCNN step is supervised and requires human labelling of initial training data. The second step of the workflow is feature engineering. Previously, we have introduced an approach where the features were image patches centered on an uniform spatial grid[48] and specific atomic location[51]. Here, we show that specific aspects of domain wall dynamics can be probed via selection of the patches centered at the specific domain walls types and introducing a time-delayed description. For example, to explore quasistatic domain morphologies we introduce descriptors centered at any domain wall. To explore the interactions between the ferroelectric and ferroelastic domain walls the patches are centered at the 180º domain walls only. Finally, to explore time dependent phenomena the patch can be centered at the wall location in the preceding time step. The third step of the workflow is unsupervised learning of low dimensional representations of the data, realized here using a multilayer rotationally-invariant autoencoder.[48] The individual elements of this workflow are discussed in detail below.

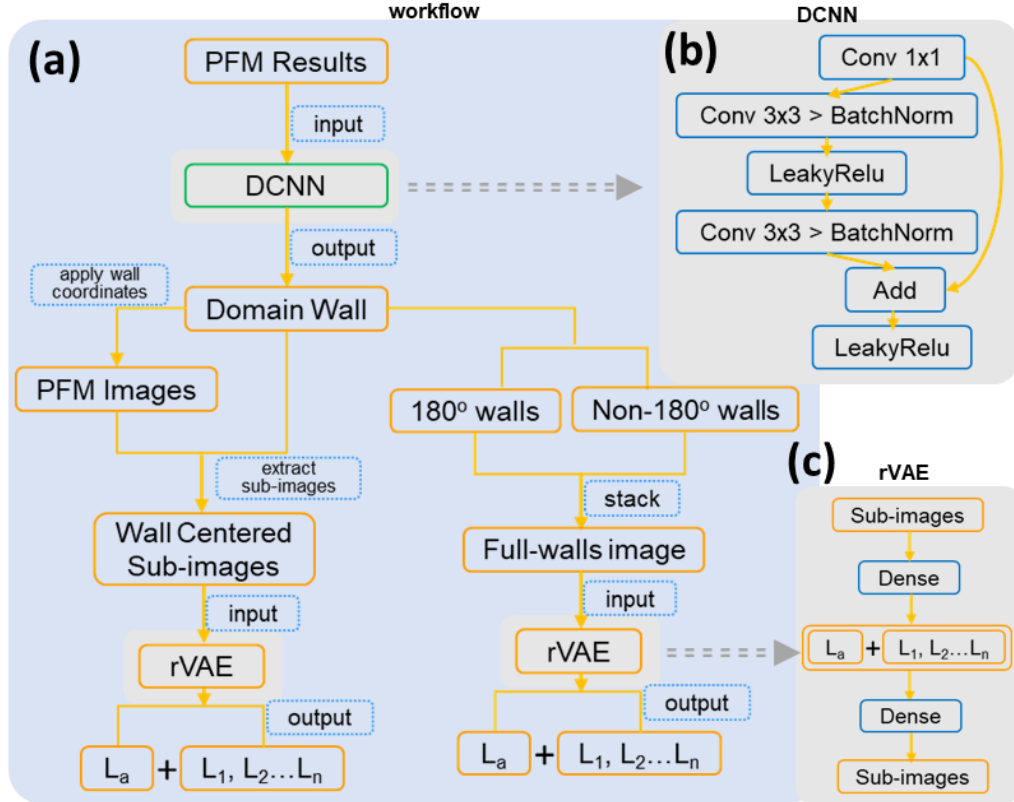

**Figure 1.** Proposed (**a**) workflow established with (**b**) deep convolutional neural network (DCNN) and (**c**) rotationally-invariant variational autoencoder (rVAE) for studying the dynamics of ferroelectric domains. The workflow combines DCNN and rVAE techniques such that DCNN detects domain walls and rVAE analyzes domain and wall features based on the information obtained information from DCNN.

As a model system of lead zirconate titanate (PZT), the polarization dynamics was explored using Interferometric Displacement Sensor (IDS-) PFM. Shown in Supplementary Video 1-2 is a series of PFM amplitude and phase images showing the domain switch upon the application of voltage. PFM phase images only show ferroelectric 180º domains while amplitude images show both ferroelectric 180º and ferroelastic non-180º domains.

The first step of performing domain wall analysis is generating wall maps, which is a computational edge detection problem. The 180º domain wall can be detected by a simple Canny filter[52] from a PFM phase image. However, detecting non-180º walls is challenging. Non-180º walls are only visible in PFM amplitude image. This makes the wall shape and level (*e.g.*, contrast on two sides of walls) complicated and hence increases the difficulty of detecting walls from amplitude images. Therefore, a deep convolutional neural network (DCNN) will be used to detect domain walls from amplitude images.

The edge detection problem is a semantic segmentation problem in that we aim to categorize every pixel in the input image as either belonging to edge (class 1) or not (class 0). However, we found empirically that popular neural networks for semantic segmentation such as U-Net[53] family of models are inadequate for this task, even when accounting for the significant class imbalance in the loss function (the best accuracy measured as intersection over union (IOU) score is always below ~60%). Among the DCNN architectures designed specifically for edge detection, the holistically-nested edge detector network (HED-net)[54] stands out due to its demonstrated ability to detect edges in a variety of (natural) images. The HED-net model consists of five blocks of back-to-back convolutional layers with a max-pooling operation in-between the blocks. The outputs from the second to fifth blocks are up-sampled to the original image size via bilinear interpolation, concatenated with the output from the first block, and passed through a sigmoid activation to generate the output. Despite its relative simplicity, the HED-net model showed state-of-the-art performance for detecting edges in natural images. Nevertheless, the application of the off-the-shelf HED-net model to our experimental data led to suboptimal results (the best accuracy ~74%) as many domain walls remained undetected and those that were detected were not well-defined (see Supplementary Figure S1). To improve the results, we increased a depth of each block while using residual connections to minimize instabilities during model training and avoid the degradation problem.[55] Essentially, the convolutional layers were replaced with residual blocks[55] (Figure 1b) as individual units of our deep learning network. In addition, we limited the number of the HED-net blocks to three, as we found from the inspection of feature maps that more blocks are not helpful for learning the relevant features in our data. This resulted in a significant (~17%) improvement of the HED-net accuracy (hereafter, we refer to it as ResHED-net). The ResHED-net (Figure 2a) performs both training phase and prediction phases in an end-to-end manner, *i.e.,* in our studies, a trained HED-net takes a PFM amplitude image as input and directly outputs a domain wall map (Figure 2b).

For the model training, the training and validation datasets were prepared as follows. First, the ground truth wall images were created by labeling domain walls of raw PFM amplitude images using ImageJ.[56] Then, sub-images with defined size were randomly extracted from raw amplitude images and ground truth images. Finally, all sub-images are shuffled and split into a training dataset and validation dataset. To train the ResHED-net model, we use 2800 pairs of raw and ground truth images (masks) created from seven source images, where 80 % was training data and 20 % validation data. To account for a class imbalance (number of pixels belonging to walls versus number of the overall pixels in the images) we used the focal loss function.[57] The deep learning model training was performed via the home-built AtomAI package[58] that uses on-the-fly data augmentation to prevent overfitting. The final accuracy (defined as the IoU score) of the ResHED network after 3000 training iterations is 91%. The evolution of training and validation loss is shown in Figure 2c. Both the training/validation dataset and the connections between inputs and outputs in ResHED-net model affect the performance of ResHED-net in certain cases, we have systematically explored these effects and summarized them in Supplementary Information Table S1-2 and provided corresponding Jupyter notebook.

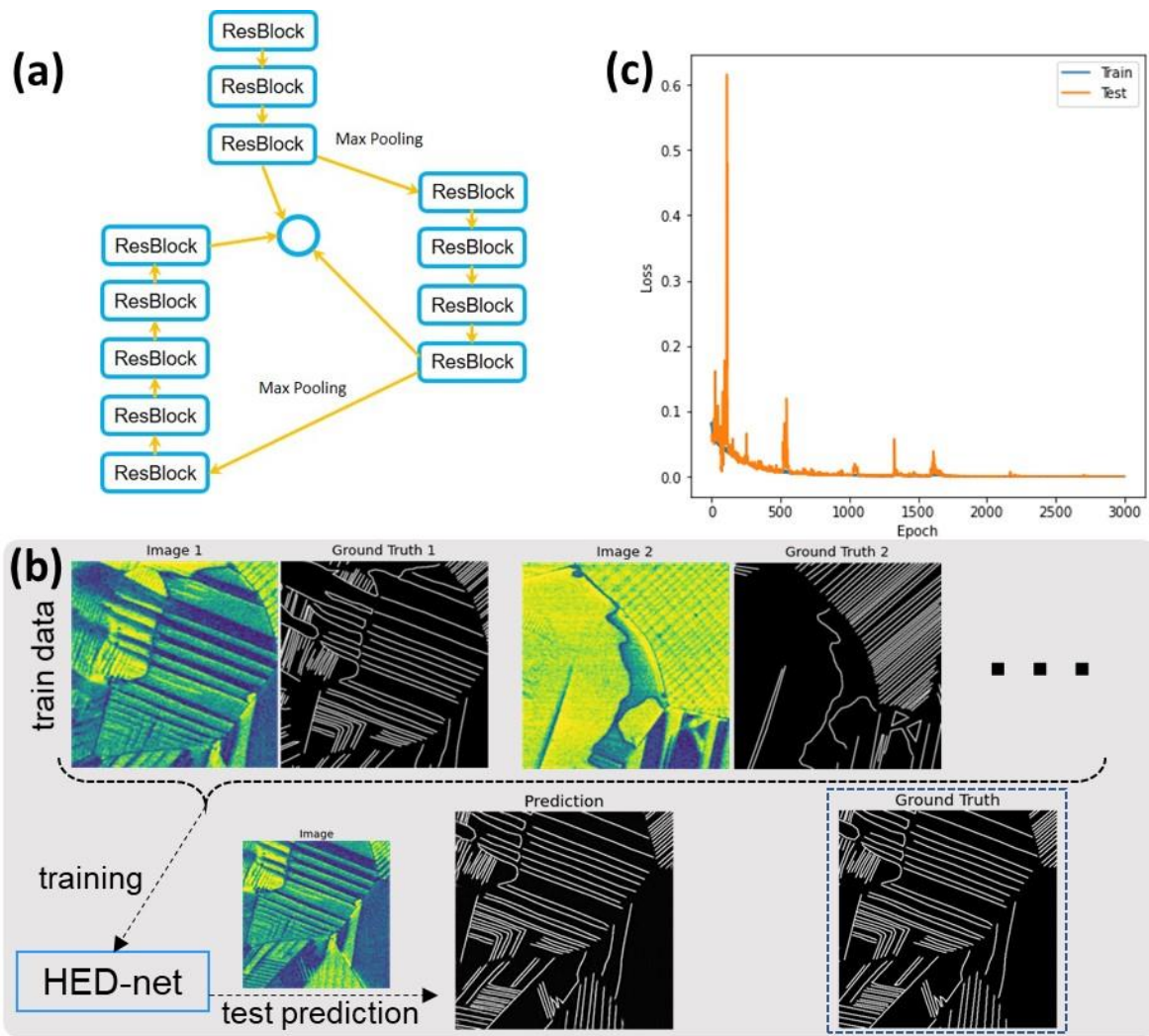

**Figure 2**. **DCNN segmentation of the PFM data**. (**a**) the architecture of ResHED network (see text for details); (**b**), ResHED-net training and test stages. In the test stage, a corresponding ground truth image is shown as a comparison with the prediction image; (**c**), the training progress expressed as the binary cross entropy loss versus the number of training iterations ("epochs") for train and validation sets.

This trained HED model generally predicts domain walls in PFM amplitude images well, with only a few minor discontinuities in non-180º wall regions (as shown in Supplementary Figure S2a-b), corresponding to the regions where the algorithm failed. However, this discontinuity problem can be resolved by denoising the raw PFM amplitude images (Supplementary Figure S2c-d). Similarly, the raw phase images were also denoised before feeding to a Canny filter in order to improve the quality of wall maps (Supplementary Figure S3). All wall maps from amplitude and phase images can be seen in Supplementary Video 3-4, respectively. Figure 3 shows a representative frame of phase and amplitude images, as well as corresponding wall maps detected by a Canny filter and HED-net, respectively. In zoom-in images, the change of 180º and non-180º walls over several frames are shown, indicating the different dynamics of 180º and non-180º walls during the application of voltage. Note that full dynamics of non-180º and 180º walls can be seen from Supplementary Video 5-6, respectively. We observe that the non-180º walls are pinned during voltage poling and hence not moving. Differently, 180º walls move upon the application of voltage. During the motion of 180º walls, not only the position but also the local curvature of 180º walls change. This domain wall motion is related to the switching of 180º domains; therefore we can also expect a change of phase contrast across 180º walls. Accordingly, the wall shape and the contrast on two sides of the wall will be taken into account in our analysis. Noteworthily, not all 180º walls show the same dynamics. For example, in Supplementary Video 6, we can observe that the 180º walls at the center region move faster that those near the edge, which is an interesting aspect about the pinning mechanism worth exploring.

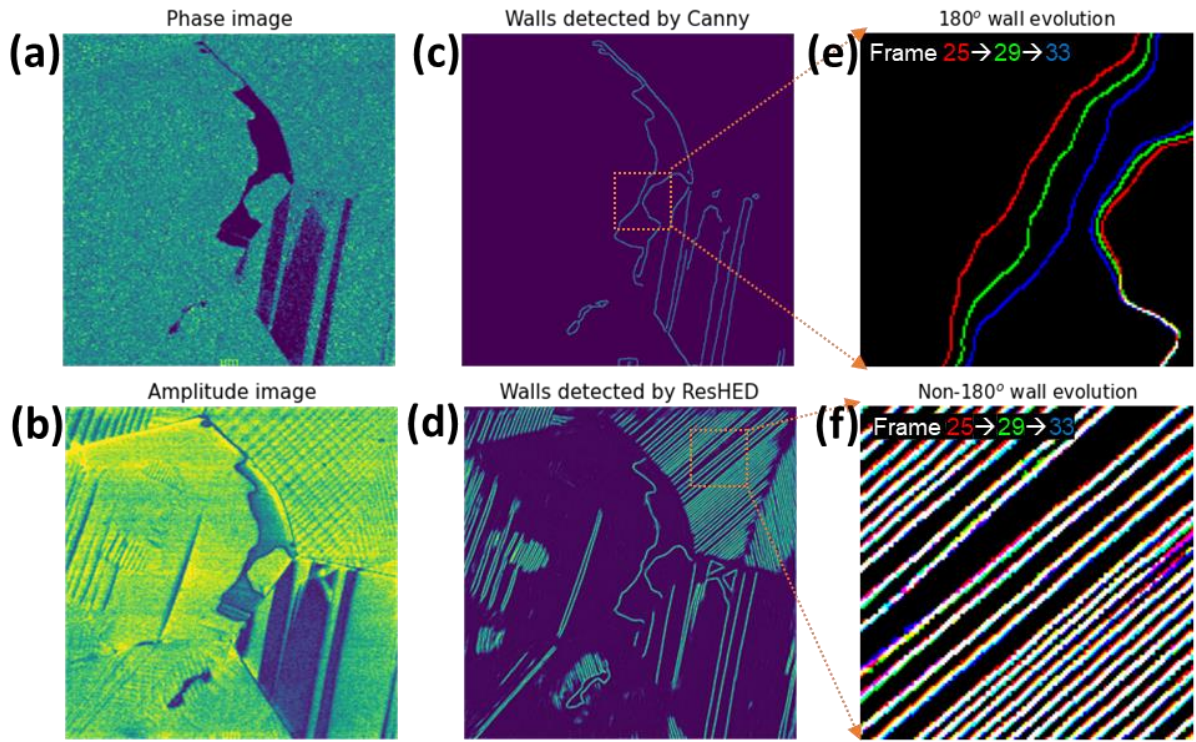

**Figure 3. Segmentation and time dynamics on PFM data.** (**a**), (**b**), raw PFM phase and amplitude images, respectively; (**c), (d),** corresponding domain wall maps predicted by Canny filter and ResHED-net, respectively; (**e), (f),** 180º and non-180º wall changes over several frames, respectively, where 180º walls move upon the application of voltage but non-180º walls are stable. Obviously, domain wall evolution involves changes in wall density and wall shape, so corresponding descriptors need to be defined in the wall dynamics analyses. Full time dynamics of 180º walls and non-180º walls can be seen in Supplementary Video 5 and 6, respectively.

To explore the domain wall dynamics, we need to define the relevant features to be used in the machine learning analysis. Typically in image recognition tasks these are either the full images, or sub-images centered on uniformly spaced grid[51], somewhat similar to the kernel filters of classical deep convolutional networks. However, it is well known that the feature selection is an effective way to encode the physical constraints in the system and directly relate the network outputs to physical mechanisms. Here, we introduce time delayed features. In these, the position of the domain wall is determined at time *t*. A number of the points is sampled from the wall position, and the sub-image centered at the selected location is sampled at time *t* and *t+dt* or *t-dt*, where *dt* is the time delay. In this fashion, a comparison of domain and wall conditions at time *t* and *t+dt* is involved in the sub-image datasets, so the information on wall dynamics during the switching was encoded. In this work, we selected *dt* = 1, so the datasets of *+dt* and *-dt* correspond to the forward dynamics and the reverse dynamics, respectively. Shown in Figure 4a are the sub-images for *t-dt*, *t*, and *t+dt*, respectively, which show the relative difference of sub-images compared to the reference wall position. In this manner, we will be able to get the information originating from the time dynamics straightforwardly by comparing the results of the analysis for forward and reverse dynamics.

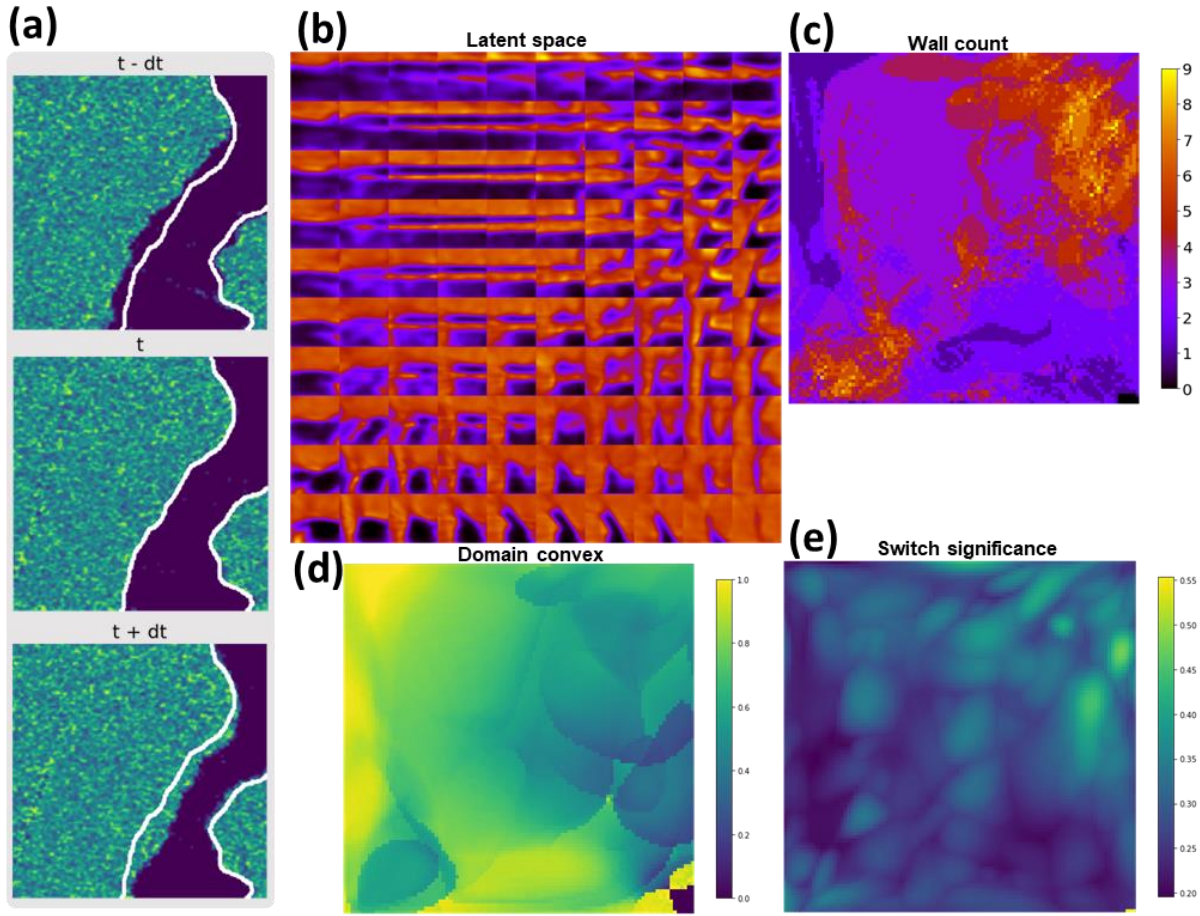

**Figure 4.** (**a**), example sub-images for *t-dt*, *t*, *t+dt*, respectively, which show the relative difference of these sub-images are clear compared to the reference wall; in these example images, *dt* is set as 3 in order to show the distinct difference; however, in rVAE analyses, *dt* is set as 1 in order to involve more detailed dynamics. **(b-e)**, rVAE analysis results of forward switch (*t vs t + dt*), **(b)**, latent space; **(c)**, domain wall count; **(d)**, domain convex; **(e)**, switch significance corresponding to latent spaces. Analogous rVAE analysis of reverse switch (*t vs t-dt*) is shown in Supplementary Figure S4.

To learn the relevant physics of the polarization switching from our experimental observations, we used the approach based on rotationally invariant variational autoencoder (r-VAE). Generally, autoencoders are a broad class of the ML algorithms that compress the input data to a small number of latent variables and subsequently aim to reconstruct the original data set. In this process, the network seeks optimal representation of the data via continuous latent variables. In variational autoencoders, the latent space is defined as a prior distribution from which the decoder samples the latent variable. The unique feature of VAEs is the capability of disentangling the representations of the data.[59, 60] While the rigorous mathematical definition of the disentanglement is still lacking, practically it manifests in latent factors describing specific aspects of data variability. For example, for hand-written digits these are the slant and width of the writing style. In certain model cases, the latent representations of the data will coincide to the ground truth factors of physical variability in the synthetic data set. However, in general the presence of such relationships has to be established for each particular case.

Here, we use the VAEs modified to allow rotationally-invariant data representations.[61, 62] In this case, rotation angle is separated as a special latent variable, and remaining variability is encoded via normal latent vectors. This is necessary since in the presence of multiple rotated versions of the same object, the rotation angle becomes the primary variability factor for multiple latent variables, precluding the disentanglement of relevant representations. Note that this limitation is also common for other linear and non-linear dimensionality reduction methods including principal component analysis, non-negative matrix factorization, and Gaussian Mixture Modeling,[63] and was the primary motivation for introduction of the rVAEs. It is also important to note that generally this behavior is related to the topological structure of the latent space of VAE, with the rotationally-invariant autoencoder being isomorphic to the SO(2) Lie group. However, VAEs can be generalized to arbitrary topology of the latent space, and hence can be used as a method for topological discovery. Previously, we demonstrated this approach for systems with evolution of structural units in 2D materials imaged via scanning transmission electron microscopy[62] and for the emergence of order in systems of rod-like nanoparticles.[50] Here, we extend this approach for probing time dependent dynamics and incorporate multilayer rVAE as a natural counterpart to multiclass DCNN to disentangle the domain wall interaction mechanisms such as pinning.

The sub-image datasets (*t vs t+dt* or *t vs t-dt*, examples shown in Figure 4a) are used as inputs to train the rVAE, the detailed training process can is shown in the provided Jupyter notebooks (see Methods for the link of Jupyter notebooks). The learned latent manifold of the rVAE (of the sub-image dataset of *t vs t+dt*) is shown in Figure 4b. To construct this image, the latent coordinates are sampled from the uniform grid of values, $L_1$ and $L_2$, and are used as an input to the trained rVAE's decoder to generate patches corresponding to each pair of the latent coordinates. Note the rich structure of the latent space, with clearly identifiable trends along primary directions. The latent space (Figure 4b) of forward switch clearly disentangles the switching behavior (top left to bottom right), domain structure (top to bottom), and domain wall complexity. Note that the continuous latent representations of the domain structures encode the physical mechanisms in the form of continuous representations of observed behaviors. Generally, higher grid density allows for more detailed information on the variability of behaviors encoded via the latent variables.

It is further illuminating to establish the relationship between the latent variables and physics-based descriptors, such as domain wall count (Figure 4c), domain convex (Figure 4d), and switch significance (Figure 4e). Here, domain wall count defines the number of continuous walls in each sub-image in the latent space. The domain wall count is derived by applying a Canny filter on each sub-image to detect edges, then counting the number of continuous walls (edges). In the physical mechanism, domain wall count can be related to the wall motion and the surrounding or the composition of the selected domain wall. First, domain wall motion changes the distance between two (or more) walls, resulting in a change in the total number of walls in a certain region. In this case, the domain wall count can be related to wall motion and switching condition. Second, the density of 180º and non-180º walls can differ in a material, thus, the wall count can also be related to the composition of domain walls in a certain region. The domain convex is defined as: $D_c = A_d/A_p$, where $D_c$ is the domain convex, $A_d$ is the area of yellow domain (can also be blue domain depending on researcher's preference), and $A_p$ is the area of the smallest convex polygon that can involve all of the yellow domain. Accordingly, the domain convex is related to the shape of the domain wall; a wall with larger curvature will have a larger value of domain convex (if the wall convex on the blue domain side). The switch significance is the maximum difference between the yellow domain and the blue domain, determined as $S_s = Max_y - Min_b$, where the $S_s$ is the switch significance, $Max_y$ is the maximum value in yellow domain, and $Min_b$ is the minimum value in the blue domain. Here the switch significance corresponds to the phase contrast on the two sides of domain walls, relevant to domain switching condition.

Altogether, these physics-based descriptors can represent the domain wall shape and switch condition. Shown in Figure 4c-e is how the domain wall counts, domain shape, and switch condition distributes in the latent space in forward switch, *i.e., t vs t + dt, dt = 1*. The same analysis was also applied in reverse switch (*i.e., t vs t + dt, dt = -1*), as shown in Supplementary Figure S4. While the overall features in latent space of the forward and reverse switch are similar, they show a different distribution. The similar features are due to using the same raw data source for constructing the forward switch and reverse switch datasets. The different distribution indicates that the time direction (forward *vs* reverse) indeed matters in rVAE analysis, suggesting the feasibility of using rVAE to extract information regarding time dynamics of domain switching and wall motion.

These analyses establish direct relationship between the domain wall dynamics and latent variables. Again, we highlight that the natural descriptors of domain dynamics are latent variables; the physical descriptors are their interpretation in terms of human-based definitions. To visualize this behavior, we further analyzed the domain wall evolution. Shown in Figure 5a-b is the sequence experimental images of a 180º domain wall showing the evolution of the wall as a function of time. Figure 5b shows the evolution of walls from frame 19 to frame 36 (corresponding video is in Supplementary Video 7), clearly indicating that the wall evolution includes not only a wall motion but also a change in wall shape. Note that, latent space indicates that the wall motion and shape change are connected with latent variables, motivating us to check the relationship between latent variables and wall dynamics. Therefore, we colored the wall evolution with latent variables in both the forward switch and reverse switch, as shown in Figure 5c and 5d, respectively. Interestingly, a comparison between the forward and reverse switch implies that latent variable 1 mostly represents the direction of wall motion (move up *vs* move down) and latent variable 2 indicates the wall curvature changes. As shown in Figure 5c-d, although the wall dynamics are opposite, moving-up walls show dark color (Region 1 in Figure 5c and Region 2 in Figure 5d) and moving-down walls show bright color (Region 2 in Figure 5c and Region 1 in Figure 5d) in latent variable 1. Meanwhile, when the curvature becomes smaller in the forward switch, the latent variable 2 is brighter (Region 3 in Figure 5c); in contrast, latent variable 2 is relatively darker when the wall curvature becomes larger (Region 3 in Figure 5d). Accordingly, we also offer two Supplementary Videos 8-11 respectively corresponding to the domain wall evolution with latent variables colors for the forward switch and reverse switch, in which the observation of the correlation between wall evolution and latent variables is more straightforward. These pieces of information will be helpful for us to check tiny changes of domain wall during a switch via looking at latent variables.

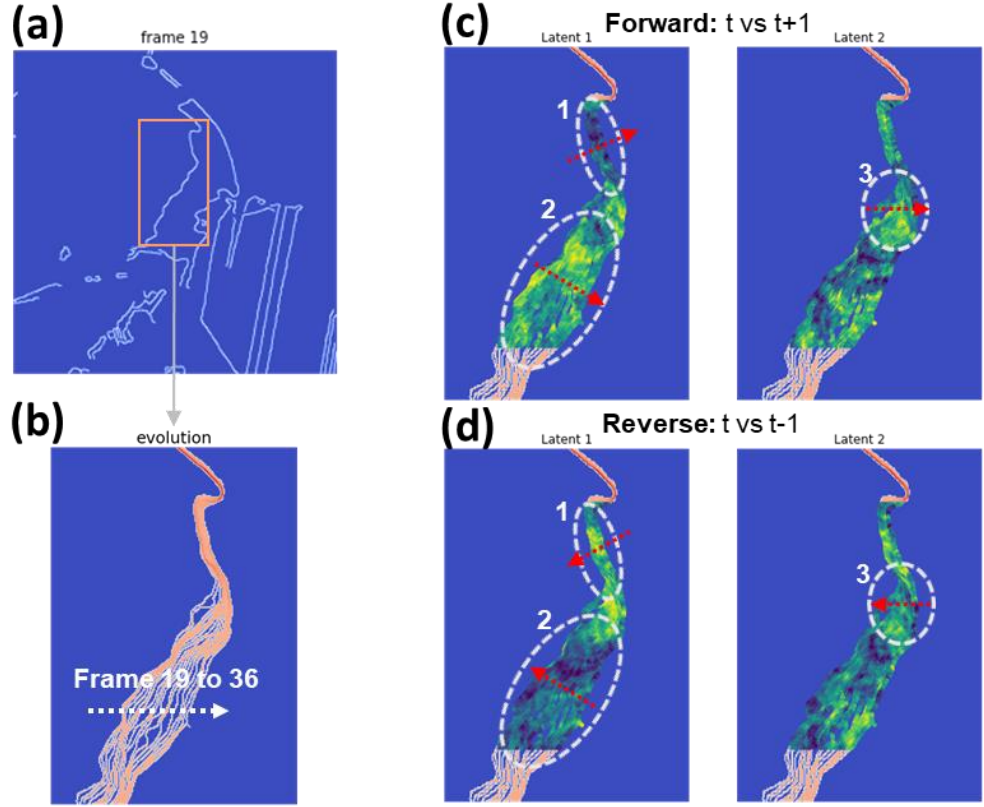

**Figure 5**. **(a), (b),** Evolution of a continuous ferroelectric domain wall between time 127 mins to 225 mins during switching, corresponding to frame 19 to 36. **(c), (d),** wall evolution colored by latent variables from forward switch and reverse switch. A video corresponding to the wall evolution in (**b**) is shown in Supplementary Video 7.

The capability of rVAE for encoding wall dynamics allows us to explore the pinning mechanism of ferroelectric walls. Owing to the interconnection of non-180º wall with ferroelasticity, it has different dynamics compared to ferroelectric 180º walls, as seen Figure 3e-f, as well as in Supplementary Video 5-6. Therefore, we take into account the effect of the interaction between ferroelectric walls and ferroelastic walls on the pinning mechanism. A three-layer image stack including 180º walls, non-180º walls, and non-180º walls was first constructed, here we chose double non-180 wall for convenience of visualization. Shown in Figure 6a-b is the construction of this three-layer dataset. The 180º wall and 180º&non-180º wall layers are from the Canny filter and HED-net, respectively, as illustrated before. Then, the non-180º wall layer is a subtraction of the 180º wall and 180º&non-180º wall layers. Then, the three-layer walls (Figure 6b) including comprehensive information regarding both ferroelectric 180º walls and ferroelastic non-180º walls, and their interactions are obtained by stacking corresponding layers. Finally, a three-layer sub-image dataset centered at ferroelectric 180º walls was prepared for rVAE analysis. Figure 6b shows a representative three-layer sub-image with ferroelectric 180º wall at the center; red walls in the image are ferroelectric walls and blue walls are ferroelastic walls.

Shown in Figure 6c is the learned latent manifold of rVAE analysis on the multi-layer sub-images, which indicates that the latent variable distribution is related to the pinning efficiency of ferroelectric walls and the interaction between ferroelectric and ferroelastic walls. The pinning efficiency can be seen as the intensity of red walls in the latent space, where a higher intensity indicates a low dynamic wall and high pinning efficiency. The interaction between ferroelectric and ferroelastic walls needs to be understood via two aspects. First is the coincidence of locations of ferroelectric and ferroelastic walls. For example, most ferroelectric (red) and ferroelastic (blue) wall locations are very close in the top-right corner of the latent space. Second is the asymmetric distribution of ferroelastic walls on two sides of the ferroelectric walls. For example, the intensity of blue color is much stronger on one side of the red wall than the other side in the bottom-left corner of the latent space. Here the intensity of the blue color means ferroelastic walls perpendicular to the ferroelectric walls and they become smeared because the rVAE does not have degrees of freedom to reconstruct them well in this case.

Accordingly, we define physical-relevant descriptors to study the pinning mechanism. The pinning efficiency is qualitatively defined as the intensity of ferroelectric walls (red). Since ferroelectric walls is constructed in the first layer, the pinning efficiency is: $E_{pin} = I_{max}^{1st} - I_{min}^{1st}$, where $E_{pin}$ is pinning efficiency, $I_{max}^{1st}$ and $I_{min}^{1st}$ are maximum and minimum intensity of the first layer, their subtraction will be the absolute intensity of ferroelectric walls (red wall). The wall coincidence is defined as the average distance between ferroelectric and ferroelastic walls. The asymmetric gradient is defined as the difference in the average intensities of ferroelastic walls on the two sides of the ferroelectric walls. This asymmetric gradient represents the asymmetric distribution of ferroelastic walls on two sides of ferroelectric walls. Such ferroelastic wall distribution can be related to local strain conditions and hence the pinning mechanism related to ferroelastic wall distribution can potentially be related to strain-mediated pinning. Note that in the whole latent space, the pinning efficiency is simultaneously related to both wall coincidence and asymmetric gradient. However, in the top-right corner of latent space, the asymmetric gradient is negligible and hence the pinning efficiency is dominantly related to wall coincidence. Shown in Figure 6d is the corresponding distribution maps of pinning efficiency and wall coincidence, where we observe similar distribution maps for pinning efficiency and wall coincidence. This confirms the correlation between ferroelectric pinning and its relative location to the ferroelastic walls. Also, in the bottom-right corner of latent space, the wall coincidence is negligible, and the pinning efficiency is now dominantly related to the asymmetric gradient. Shown in Figure 6e is the distribution maps of the pinning efficiency and the asymmetric gradient, in which we observe consistent distributions in two maps. This also confirms the correlation between the ferroelectric pinning and the distribution of ferroelastic walls in near regions.

Further by distributing latent variables in the three-layer wall image (Figure 6f-g), we found that latent variables represent the contribution of wall coincidence and asymmetric gradient to ferroelectric pinning, respectively. Shown in Figure 6f-h is the distribution of encoded angle and latent variables, the color scale is based on the values of latent variables. The bright color means strong contribution to pinning efficiency and dark color means weak contribution. In Figure 6g, we observe that the bright color mainly distributed at the bottom-right corner, this is because of the consistent locations of ferroelectric and ferroelastic walls in this region (which can be seen in Figure 3c-d). In Figure 6h, the bright color distribution is also consistent with the asymmetric distribution of ferroelastic non-180º walls, suggesting the pinning is governed by asymmetric distribution of ferroelastic walls in this region. Above all, we can find that the rVAE latent variables are physically-relevant variables. Latent variable 1 indicates the contribution of the wall coincidence to the pinning efficiency and latent variable 2 indicates the contribution of the asymmetric distributed ferroelastic walls to the pinning efficiency.

Note that the sub-image size for above three-layer analysis is 30 x 30 pixels, while any size of sub-image is doable through the python notebook. In our analysis, we found that the latent space becomes collapsed when the sub-image is larger than 60 pixels. In addition, the time dynamics (*e.g.*, *t vs t+dt*) can also be included in the three-layer rVAE analysis. We provide a result for a three-layer rVAE analysis for *t vs t+1* in Supplementary Figure S5. In this case, the latent space shows more ferroelectric walls (red walls) with low pinning efficiency because of the time dynamic factor in the analysis. Nonetheless, the latent variables colored walls still show the contributions of the wall coincidence and the asymmetric gradient to the ferroelectric pinning.

To summarize, we applied machine learning to analyze PFM results of domain wall dynamics. A Canny filter and DCNN were utilized to detect 180º and non180º domain walls based on the PFM phase and amplitude images, respectively. Then, sub-images centered at 180º domain walls were used for rVAE analysis to encode information of domain switch and wall dynamics. The same PFM result was used to construct two sub-image datasets representing forward domain switch and reverse domain switch, respectively. The time dynamics of the forward and reverse switch were encoded into latent variables, where we found that the latent variables can represent the moving direction and the curvature change of domain walls. Finally, through constructing a three-layer dataset including both the ferroelectric 180º wall layer and ferroelastic non-180º wall layer, we investigated the connection of 180º wall dynamics to the interaction between ferroelectric and ferroelastic walls. The rVAE discovered two factors affecting the ferroelectric wall dynamics (pinning mechanism), which are coincidence locations of ferroelectric and ferroelastic walls and the asymmetric distribution of ferroelastic walls around the ferroelectric walls. Finally, we believe that the approach developed in this work will be universally useful for time-dependent dynamics of complex materials.

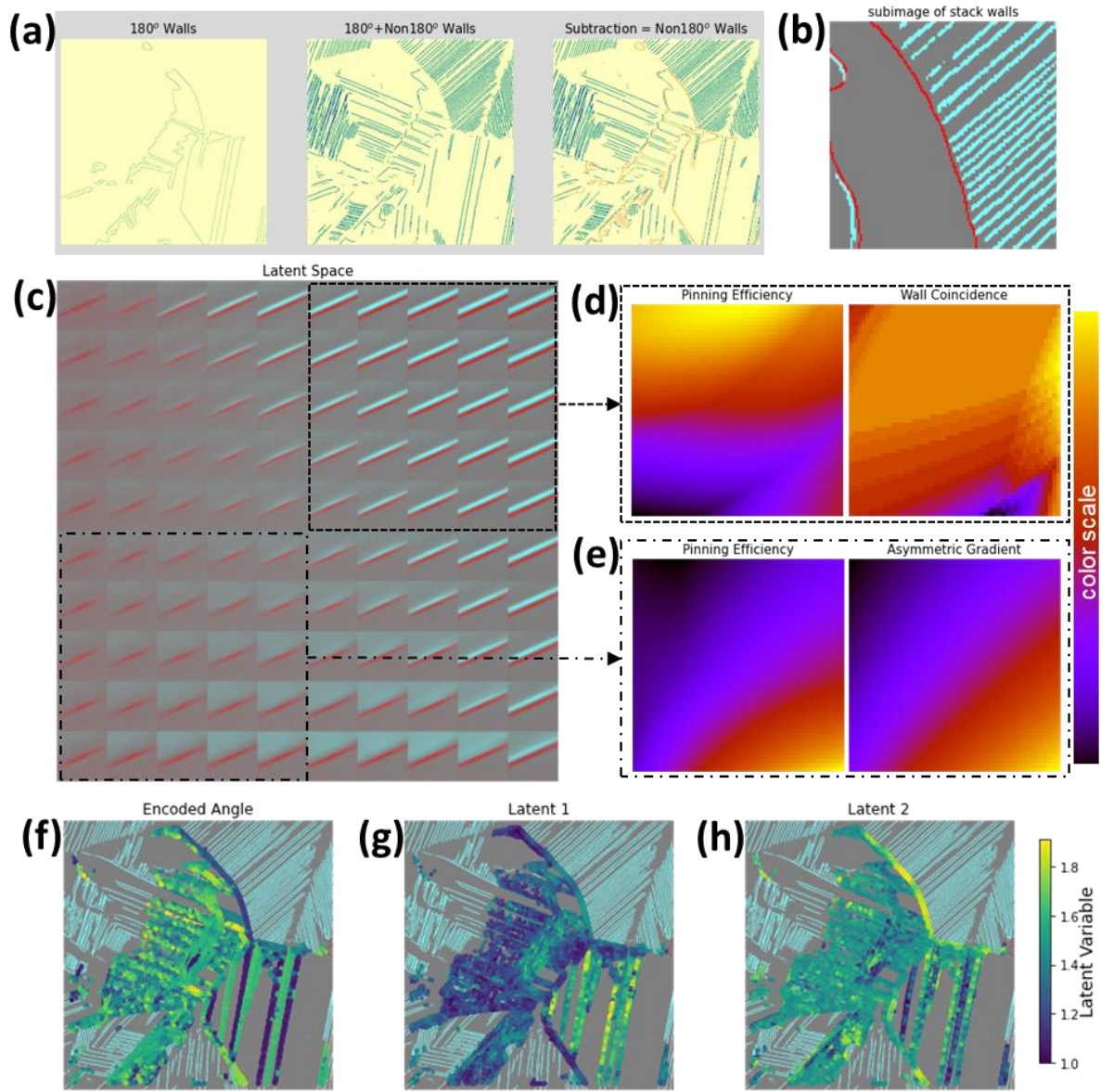

**Figure 6.** (**a-b**), construction of the three-layer dataset containing comprehensive information of 180°wall and non-180° wall; (**a**), 180° ferroelectric walls, 180° and non-180° walls, and non-180° ferroelastic walls, (**b**) a representative three-layer sub-image showing 180° (red) and non-180° walls (blue). (**c**), latent space; this latent space shows the interaction of ferroelectric 180° walls (red) and ferroelastic non-180° walls (blue), which reflects the pinning mechanism of ferroelectric walls by ferroelasticity. (**d**), these plots correspond to top-right of latent space, which shows the correlation of pinning efficiency of ferroelectric walls and the coincidence of ferroelectric and ferroelastic walls. (**e**), these plots correspond to bottom-left of latent space, which shows the correlation of pinning efficiency of ferroelectric walls and the asymmetric gradient aside of ferroelastic wall intensity aside ferroelectric walls. (**f-h**), encoded angle and latent variables distribution in the image integrated over full three-layer dataset.

## Methods

*Data analysis*

The detailed methodologies of DCNN and rVAE analysis is established in Jupyter notebooks and are available from https://git.io/J3Cvl.

*PZT sample and PFM imaging*

The measured sample was a polycrystalline lead zirconate titanate - PbZr1−x Ti x O3 (PZT) with x~0.4 from Morgan Advanced Materials. The samples were cut with a diamond band saw and polished first with diamond hand polishing pads (from 50-3000 grit) and then with successively finer diamond polishing compounds, finishing with 50,000 grit lapidary paste. These materials were overstock for tube scanners in Atomic Force Microscope. The PFM imaging was performed using a commercial Cypher AFM system (Asylum Research, Santa Barbara, CA).

**Conflict of Interest**

The authors declare no conflict of interest.

**Authors Contribution**

S.V.K. conceived the project and M.Z. realized the customized HED-net and rVAE via AtomAI package. Y.L. performed detailed analyses with basic workflow from S.V.K. T.W. and Y.L. adapted the HED-net for domain wall detection. R.P. performed PFM measurements of PZT samples. S.V.K. and Y.L. wrote the manuscript. All authors contributed to discussions and the final manuscript.

**Acknowledgements**

This effort (ML) is based upon work supported by the U.S. Department of Energy, Office of Science, Office of Basic Energy Sciences Energy Frontier Research Centers program under Award Number DE-SC0021118 (Y.L., S.V.K.), and the Oak Ridge National Laboratory's Center for Nanophase Materials Sciences (CNMS), a U.S. Department of Energy, Office of Science User Facility (M.Z.).

**Data Availability Statement**

The data that support the findings of this study are available at https://git.io/J3Cvl.

# References

1. A.K. Tagantsev, L. E. C., and J. Fousek, *Domains in Ferroic Crystals and Thin Films*. Springer: New York, 2010.
2. Waser, R.; Aono, M., Nanoionics-based resistive switching memories. *Nat. Mater.* **2007,** *6* (11), 833-840.
3. Waser, R., Nanoelectronics and Information Technology *Nanoelectronics and Information Technology* **2012**.
4. Damjanovic, D., Ferroelectric, dielectric and piezoelectric properties of ferroelectric thin films and ceramics. *Rep. Prog. Phys.* **1998,** *61* (9), 1267-1324.
5. Setter, N.; Damjanovic, D.; Eng, L.; Fox, G.; Gevorgian, S.; Hong, S.; Kingon, A.; Kohlstedt, H.; Park, N. Y.; Stephenson, G. B.; Stolitchnov, I.; Tagantsev, A. K.; Taylor, D. V.; Yamada, T.; Streiffer, S., Ferroelectric thin films: Review of materials, properties, and applications. *J. Appl. Phys.* **2006,** *100* (5).
6. Shin, Y.-H.; Grinberg, I.; Chen, I.-W.; Rappe, A. M., Nucleation and growth mechanism of ferroelectric domain-wall motion. *Nature* **2007,** *449* (7164), 881-884.
7. McGilly, L.; Yudin, P.; Feigl, L.; Tagantsev, A.; Setter, N., Controlling domain wall motion in ferroelectric thin films. *Nature nanotechnology* **2015,** *10* (2), 145-150.
8. Jesse, S.; Rodriguez, B. J.; Choudhury, S.; Baddorf, A. P.; Vrejoiu, I.; Hesse, D.; Alexe, M.; Eliseev, E. A.; Morozovska, A. N.; Zhang, J., Direct imaging of the spatial and energy distribution of nucleation centres in ferroelectric materials. *Nature materials* **2008,** *7* (3), 209-215.
9. Maksymovych, P.; Jesse, S.; Huijben, M.; Ramesh, R.; Morozovska, A.; Choudhury, S.; Chen, L.-Q.; Baddorf, A. P.; Kalinin, S. V., Intrinsic nucleation mechanism and disorder effects in polarization switching on ferroelectric surfaces. *Physical review letters* **2009,** *102* (1), 017601.
10. Marincel, D. M.; Zhang, H.; Britson, J.; Belianinov, A.; Jesse, S.; Kalinin, S. V.; Chen, L.; Rainforth, W. M.; Reaney, I. M.; Randall, C. A., Domain pinning near a single-grain boundary in tetragonal and rhombohedral lead zirconate titanate films. *Physical Review B* **2015,** *91* (13), 134113.
11. Marincel, D. M.; Zhang, H.; Jesse, S.; Belianinov, A.; Okatan, M. B.; Kalinin, S. V.; Rainforth, W. M.; Reaney, I. M.; Randall, C. A.; Trolier-McKinstry, S., Domain wall motion across various grain boundaries in ferroelectric thin films. *Journal of the American Ceramic Society* **2015,** *98* (6), 1848-1857.
12. Rojac, T.; Kosec, M.; Budic, B.; Setter, N.; Damjanovic, D., Strong ferroelectric domain-wall pinning in BiFeO 3 ceramics. *Journal of Applied Physics* **2010,** *108* (7), 074107.
13. Rodriguez, B. J.; Chu, Y.; Ramesh, R.; Kalinin, S. V., Ferroelectric domain wall pinning at a bicrystal grain boundary in bismuth ferrite. *Applied Physics Letters* **2008,** *93* (14), 142901.
14. Kalinin, S. V.; Morozovska, A. N.; Chen, L. Q.; Rodriguez, B. J., Local polarization dynamics in ferroelectric materials. *Reports on Progress in Physics* **2010,** *73* (5), 056502.
15. Damjanovic, D.; Demartin, M., Contribution of the irreversible displacement of domain walls to the piezoelectric effect in barium titanate and lead zirconate titanate ceramics. *Journal of Physics: Condensed Matter* **1997,** *9* (23), 4943.
16. Seshadri, S. B.; Prewitt, A. D.; Studer, A. J.; Damjanovic, D.; Jones, J. L., An in situ diffraction study of domain wall motion contributions to the frequency dispersion of the piezoelectric coefficient in lead zirconate titanate. *Applied Physics Letters* **2013,** *102* (4), 042911.
17. Tutuncu, G.; Li, B.; Bowman, K.; Jones, J. L., Domain wall motion and electromechanical strain in lead-free piezoelectrics: Insight from the model system (1− x) Ba (Zr0. 2Ti0. 8) O3–x (Ba0. 7Ca0. 3) TiO3 using in situ high-energy X-ray diffraction during application of electric fields. *Journal of Applied Physics* **2014,** *115* (14), 144104.
18. Glaum, J.; Granzow, T.; Rödel, J., Evaluation of domain wall motion in bipolar fatigued lead-zirconate-titanate: A study on reversible and irreversible contributions. *Journal of Applied Physics* **2010,** *107* (10), 104119.

19. Jones, J. L.; Slamovich, E. B.; Bowman, K. J., Domain texture distributions in tetragonal lead zirconate titanate by x-ray and neutron diffraction. *J. Appl. Phys.* **2005,** *97* (3), 6.
20. Jones, J. L.; Hoffman, M.; Daniels, J. E.; Studer, A. J., Direct measurement of the domain switching contribution to the dynamic piezoelectric response in ferroelectric ceramics. *Appl. Phys. Lett.* **2006,** *89* (9), 3.
21. Ishibashi, Y.; Iwata, M.; Salje, E., Polarization reversals in the presence of 90 degrees domain walls. *Japanese Journal of Applied Physics Part 1-Regular Papers Brief Communications & Review Papers* **2005,** *44* (10), 7512-7517.
22. Eliseev, E. A.; Morozovska, A. N.; Nelson, C. T.; Kalinin, S. V., Intrinsic structural instabilities of domain walls driven by gradient coupling: Meandering antiferrodistortive-ferroelectric domain walls in BiFe O 3. *Physical Review B* **2019,** *99* (1), 014112.
23. Maksymovych, P.; Seidel, J.; Chu, Y. H.; Wu, P.; Baddorf, A. P.; Chen, L.-Q.; Kalinin, S. V.; Ramesh, R., Dynamic conductivity of ferroelectric domain walls in BiFeO3. *Nano letters* **2011,** *11* (5), 1906-1912.
24. Yang, S. M.; Morozovska, A. N.; Kumar, R.; Eliseev, E. A.; Cao, Y.; Mazet, L.; Balke, N.; Jesse, S.; Vasudevan, R. K.; Dubourdieu, C., Mixed electrochemical–ferroelectric states in nanoscale ferroelectrics. *Nature Physics* **2017,** *13* (8), 812-818.
25. Eliseev, E. A.; Morozovska, A. N.; Kalinin, S. V.; Li, Y.; Shen, J.; Glinchuk, M. D.; Chen, L.-Q.; Gopalan, V., Surface effect on domain wall width in ferroelectrics. *Journal of Applied Physics* **2009,** *106* (8), 084102.
26. Kolosov, O.; Gruverman, A.; Hatano, J.; Takahashi, K.; Tokumoto, H., NANOSCALE VISUALIZATION AND CONTROL OF FERROELECTRIC DOMAINS BY ATOMIC-FORCE MICROSCOPY. *Phys. Rev. Lett.* **1995,** *74* (21), 4309-4312.
27. Gruverman, A. L.; Hatano, J.; Tokumoto, H., Scanning force microscopy studies of domain structure in BaTiO3 single crystals. *Japanese Journal of Applied Physics Part 1-Regular Papers Short Notes & Review Papers* **1997,** *36* (4A), 2207-2211.
28. Gruverman, A.; Auciello, O.; Ramesh, R.; Tokumoto, H., Scanning force microscopy of domain structure in ferroelectric thin films: imaging and control. *Nanotechnology* **1997,** *8*, A38-A43.
29. Franke, K.; Besold, J.; Haessler, W.; Seegebarth, C., MODIFICATION AND DETECTION OF DOMAINS ON FERROELECTRIC PZT FILMS BY SCANNING FORCE MICROSCOPY. *Surface Science* **1994,** *302* (1-2), L283-L288.
30. Franke, K.; Weihnacht, M., EVALUATION OF ELECTRICALLY POLAR SUBSTANCES BY ELECTRIC SCANNING FORCE MICROSCOPY .1. MEASUREMENT SIGNALS DUE TO MAXWELL STRESS. *Ferroelectr. Lett. Sect.* **1995,** *19* (1-2), 25-33.
31. Takata, K.; Kushida, K.; Torii, K., STRAIN-IMAGING OBSERVATION OF PB(ZR,TI)O-3 THIN-FILMS. *Japanese Journal of Applied Physics Part 1-Regular Papers Short Notes & Review Papers* **1995,** *34* (5B), 2890-2894.
32. Balke, N.; Gajek, M.; Tagantsev, A. K.; Martin, L. W.; Chu, Y. H.; Ramesh, R.; Kalinin, S. V., Direct Observation of Capacitor Switching Using Planar Electrodes. *Adv. Funct. Mater.* **2010,** *20* (20), 3466-3475.
33. Dehoff, C.; Rodriguez, B. J.; Kingon, A. I.; Nemanich, R. J.; Gruverman, A.; Cross, J. S., Atomic force microscopy-based experimental setup for studying domain switching dynamics in ferroelectric capacitors. *Rev. Sci. Instrum.* **2005,** *76* (2).
34. Kim, D. J.; Jo, J. Y.; Kim, T. H.; Yang, S. M.; Chen, B.; Kim, Y. S.; Noh, T. W., Observation of inhomogeneous domain nucleation in epitaxial Pb(Zr,Ti)O(3) capacitors. *Appl. Phys. Lett.* **2007,** *91* (13).
35. Alexe, M.; Gruverman, A.; Harnagea, C.; Zakharov, N. D.; Pignolet, A.; Hesse, D.; Scott, J. F., Switching properties of self-assembled ferroelectric memory cells. *Appl. Phys. Lett.* **1999,** *75* (8), 1158-1160.

36. Jesse, S.; Rodriguez, B. J.; Choudhury, S.; Baddorf, A. P.; Vrejoiu, I.; Hesse, D.; Alexe, M.; Eliseev, E. A.; Morozovska, A. N.; Zhang, J.; Chen, L. Q.; Kalinin, S. V., Direct imaging of the spatial and energy distribution of nucleation centres in ferroelectric materials. *Nat. Mater.* **2008,** *7* (3), 209-215.
37. Gruverman, A.; Kalinin, S. V., Piezoresponse force microscopy and recent advances in nanoscale studies of ferroelectrics. *Journal of Materials Science* **2006,** *41* (1), 107-116.
38. Kalinin, S. V.; Shao, R.; Bonnell, D. A., Local phenomena in oxides by advanced scanning probe microscopy. *J. Am. Ceram. Soc.* **2005,** *88* (5), 1077-1098.
39. Jesse, S.; Lee, H. N.; Kalinin, S. V., Quantitative mapping of switching behavior in piezoresponse force microscopy. *Rev. Sci. Instrum.* **2006,** *77* (7), 073702.
40. Balke, N.; Bdikin, I.; Kalinin, S. V.; Kholkin, A. L., Electromechanical Imaging and Spectroscopy of Ferroelectric and Piezoelectric Materials: State of the Art and Prospects for the Future. *Journal of the American Ceramic Society* **2009,** *92* (8), 1629-1647.
41. Ivry, Y.; Chu, D. P.; Scott, J. F.; Salje, E. K. H.; Durkan, C., Unexpected Controllable Pair-Structure in Ferroelectric Nanodomains. *Nano Lett.* **2011,** *11* (11), 4619-4625.
42. Rodriguez, B. J.; Nemanich, R. J.; Kingon, A.; Gruverman, A.; Kalinin, S. V.; Terabe, K.; Liu, X. Y.; Kitamura, K., Domain growth kinetics in lithium niobate single crystals studied by piezoresponse force microscopy. *Appl. Phys. Lett.* **2005,** *86* (1).
43. Paruch, P.; Giamarchi, T.; Tybell, T.; Triscone, J. M., Nanoscale studies of domain wall motion in epitaxial ferroelectric thin films. *J. Appl. Phys.* **2006,** *100* (5).
44. Paruch, P.; Giamarchi, T.; Triscone, J. M., Domain wall roughness in epitaxial ferroelectric PbZr(0.2)Ti(0.8)O(3) thin films. *Phys. Rev. Lett.* **2005,** *94* (19).
45. Shishkin, E. I.; Shur, V. Y.; Schlaphof, F.; Eng, L. M., Observation and manipulation of the as-grown maze domain structure in lead germanate by scanning force microscopy. *Appl. Phys. Lett.* **2006,** *88* (25).
46. Rodriguez, B. J.; Jesse, S.; Baddorf, A. P.; Kim, S. H.; Kalinin, S. V., Controlling polarization dynamics in a liquid environment: From localized to macroscopic switching in ferroelectrics. *Phys. Rev. Lett.* **2007,** *98* (24).
47. Shur, V.; Akhmatkhanov, A.; Lobov, A.; Turygin, A., Self-assembled domain structures: From micro- to nanoscale. *Journal of Advanced Dielectrics* **2015,** *5* (2).
48. Kalinin, S. V.; Steffes, J. J.; Huey, B. D.; Ziatdinov, M., Disentangling ferroelectric domain wall geometries and pathways in dynamic piezoresponse force microscopy via unsupervised machine learning. *arXiv preprint arXiv:2007.06194* **2020**.
49. Valleti, M.; Ignatans, R.; Kalinin, S. V.; Tileli, V., Decoding the mechanisms of phase transitions from in situ microscopy observations. *arXiv preprint arXiv:2011.09513* **2020**.
50. Ignatans, R.; Ziatdinov, M.; Vasudevan, R.; Valleti, M.; Tileli, V.; Kalinin, S. V., Latent mechanisms of polarization switching from in situ electron microscopy observations. *arXiv preprint arXiv:2011.11869* **2020**.
51. Kalinin, S. V.; Dyck, O.; Ghosh, A.; Sumpter, B. G.; Ziatdinov, M., Unsupervised Machine Learning Discovery of Chemical Transformation Pathways from Atomically-Resolved Imaging Data. *arXiv preprint arXiv:2010.09196* **2020**.
52. Canny, J., A computational approach to edge detection. *IEEE Transactions on pattern analysis and machine intelligence* **1986,** (6), 679-698.
53. Ronneberger, O.; Fischer, P.; Brox, T., U-net: Convolutional networks for biomedical image segmentation. In *International Conference on Medical image computing and computer-assisted intervention*, Springer: Cham, 2015; pp 234-241.
54. Xie, S.; Tu, Z. In *Holistically-nested edge detection*, Proceedings of the IEEE international conference on computer vision, 2015; pp 1395-1403.
55. He, K.; Zhang, X.; Ren, S.; Sun, J. In *Deep residual learning for image recognition*, Proceedings of the IEEE conference on computer vision and pattern recognition, 2016; pp 770-778.

56. https://imagej.nih.gov/ij/.
57. Lin, T.-Y.; Goyal, P.; Girshick, R.; He, K.; Dollár, P. In *Focal loss for dense object detection*, Proceedings of the IEEE international conference on computer vision, 2017; pp 2980-2988.
58. **!!! INVALID CITATION !!! {}**.
59. Kingma, D. P.; Welling, M., An Introduction to Variational Autoencoders. *Foundations and Trends® in Machine Learning* **2019,** *12* (4), 307-392.
60. Watters, N.; Matthey, L.; Burgess, C. P.; Lerchner, A., Spatial broadcast decoder: A simple architecture for learning disentangled representations in VAEs. *arXiv preprint arXiv:1901.07017* **2019**.
61. Bepler, T.; Zhong, E.; Kelley, K.; Brignole, E.; Berger, B., Explicitly disentangling image content from translation and rotation with spatial-VAE. *Advances in Neural Information Processing Systems* **2019**, 15409-15419.
62. Kalinin, S. V.; Dyck, O.; Jesse, S.; Ziatdinov, M., Exploring order parameters and dynamic processes in disordered systems via variational autoencoders. *Science Advances* **2021,** *7* (17), eabd5084.
63. Ziatdinov, M.; Dyck, O.; Jesse, S.; Kalinin, S. V., Atomic Mechanisms for the Si Atom Dynamics in Graphene: Chemical Transformations at the Edge and in the Bulk. *Adv. Funct. Mater.* **2019,** *29* (52), 8.